\documentclass[12pt]{iopart}
 
\usepackage{hyperref}
\usepackage{graphicx}
\usepackage{iopams}

\expandafter\let\csname equation*\endcsname\relax

\expandafter\let\csname endequation*\endcsname\relax

\usepackage{bbold}
\usepackage{xcolor}
\usepackage[xcolor,final]{changes}

\usepackage{empheq}
\usepackage{tikz}
\usetikzlibrary{decorations.pathmorphing, positioning, shapes.geometric}

\newcommand{\ket}[1]{|#1\rangle}
\newcommand{\bra}[1]{\langle#1|}

\begin{document}

\title{Entanglement of truncated quantum states}

\author{Giacomo Sorelli$^{1,2,3}$, Vyacheslav N. Shatokhin$^1$, Filippus S. Roux$^{4}$ and Andreas Buchleitner$^{1}$} 
\address{$1$ Physikalisches Institut, Albert-Ludwigs-Universit\"at Freiburg, Hermann-Herder-Stra\ss e 3, D-79104 Freiburg, Fed. Rep. of Germany}
\address{$2$ Laboratoire Kastler Brossel, Sorbonne Universit\'e, ENS-Universit\'e PSL,
Coll\`ege de France, CNRS; 4 place Jussieu, F-75252 Paris, France}
\address{$3$ DEMR, ONERA, 6  Chemin de la Hunière, F-91123, Palaiseau,France}
\address{$4$ National Metrology Institute of South Africa, Meiring Naud\'e Road, Brummeria 0040, Pretoria, South Africa}

\begin{abstract}
We investigate the impact of Hilbert-space truncation upon the entanglement of an initially maximally entangled $m\times m$ bipartite quantum state, after propagation under an entanglement-preserving $n \times n$ ($n\geq m$) unitary.
Truncation -- physically enforced, e.g., by a detector's finite cross section -- projects the state onto an $s \times s$-dimensional subspace ($3\leq s \leq n$). For a random local unitary evolution, we obtain a simple analytical formula that expresses the truncation-induced entanglement loss as a function of $n$, $m$ and $s$.
\end{abstract}


\section{Introduction}
Entanglement is one of the defining features of quantum mechanics, and also a fundamental resource for many quantum information protocols \cite{horodecki2009}. 
Many theoretical and experimental studies were dedicated to the entanglement of a pair of two-level systems (qubits). 
Bipartite entanglement of high-dimensional (qudit) systems is less studied. 
Yet, from a fundamental point of view, a better understanding of entangled qudits could clarify some subtleties of quantum physics. 
For instance, qudits were shown to enhance non-classical effects when compared to qubits, since they allow for stronger violations of local realism \cite{KaszlikowskiPRL2000, CollinsPRL2002}. 
Moreover, from a more pragmatic point of view, high-dimensional quantum states have a higher information capacity than simple qubits and allow quantum key distribution protocols to tolerate higher noise thresholds \cite{CerfPRL2002}.

In photonic systems, (entangled) qudits are encoded in finite-dimensional subspaces of high (eventually, infinite) dimensional Hilbert spaces.
This is achieved either by using spatial modes (e.g. orbital angular momentum \cite{Mair2001,Franke-Arnold2002,Dada2011}) or by discretizing continuous degrees of freedom such as frequency \cite{REICHERT1999,Roslund2013} or time \cite{Thew:2004,Brendel:1999}.
Moreover, such initially finite-dimensional states can spread over the entire Hilbert space, in the course of their dynamical evolution.
For example, this is the case for photonic orbital-angular-momentum carrying states \cite{allen92} when transmitted through free space \cite{IbrahimPRA2013,Krenn14197,Leonhard2015,Roux2015} or across optical fibers \cite{Bozinovic1545}. 
However, the output states are often projected onto the encoding subspace, or another finite-dimensional subspace determined by the finite size and resolution of detectors \cite{ecker2019}.
\added{Such a scenario is also relevant for the emerging field of photonic quantum information in complex scattering media \cite{Defiennee1501054, PhysRevLett.121.233601, Leedumrongwatthanakun2019}, where the access to a limited number of output modes strongly influences the transmission properties of the system \cite{PhysRevLett.111.063901}.}
\replaced{In all these contexts, when}{If} states inside and outside of the encoding subspace are strongly coupled to each other, \replaced{Hilbert-space}{such }truncation leads to a decay of the output state's norm and can also affect its entanglement.
However, while it has always been clear that a truncated state must be renormalized, the influence of truncation on entanglement has never been discussed so far.

\begin{figure}[t]
\centering
\begin{tikzpicture}[scale = 0.7]
\draw[thick,black] (0,0) node[black,left] {-3}-- (1.,0);
\draw[thick,black] (0,0.5) node[black,left] {-2} -- (1.,0.5);
\draw[thick,black] (0,1) node[black,left] {-1} -- (1.,1);
\draw[thick,black] (0,1.5) node[black,left] {0} -- (1.,1.5);
\draw[thick,black] (0,2.0) node[black,left] {1} -- (1.,2.0);
\draw[thick,black] (0,2.5) node[black,left] {2} -- (1.,2.5);
\draw[thick,black] (0,3.0) node[black,left] {3} -- (1.,3.0);
\filldraw[red] (0.5,1) circle (4pt);
\filldraw[red] (0.5,2) circle (4pt);
\draw[thick,black] (1.5,0) -- (2.5,0);
\draw[thick,black] (1.5,0.5) -- (2.5,0.5);
\draw[thick,black] (1.5,1) -- (2.5,1);
\draw[thick,black] (1.5,1.5) -- (2.5,1.5);
\draw[thick,black] (1.5,2.0) -- (2.5,2.0);
\draw[thick,black] (1.5,2.5) -- (2.5,2.5);
\draw[thick,black] (1.5,3.0) -- (2.5,3.0);
\filldraw[blue] (2.,1) circle (4pt);
\filldraw[blue] (2.,2) circle (4pt);
\draw[very thick,->] (1.25,3.3) .. controls (3.25,4) .. (5.2,3.25);
\node at (3.3,4.2) {{\bf Evolution}};
\draw[thick,black] (4,0) -- (5,0);
\draw[thick,black] (4,0.5) -- (5,0.5);
\draw[thick,black] (4,1) -- (5,1);
\draw[thick,black] (4,1.5) -- (5,1.5);
\draw[thick,black] (4,2.0) -- (5,2.0);
\draw[thick,black] (4,2.5) -- (5,2.5);
\draw[thick,black] (4,3.0) -- (5,3.0);
\filldraw[red!80] (4.5,1) circle (4pt);
\filldraw[red!30] (4.5,2) circle (4pt);
\filldraw[red!60] (4.5,1.5) circle (4pt);
\filldraw[red!50] (4.5,2.5) circle (4pt);
\filldraw[red!70] (4.5,0.5) circle (4pt);
\filldraw[red!10] (4.5,3.) circle (4pt);
\filldraw[red!20] (4.5,0.) circle (4pt);
\draw[thick,black] (5.5,0) -- (6.5,0);
\draw[thick,black] (5.5,0.5) -- (6.5,0.5);
\draw[thick,black] (5.5,1) -- (6.5,1);
\draw[thick,black] (5.5,1.5) -- (6.5,1.5);
\draw[thick,black] (5.5,2.0) -- (6.5,2.0);
\draw[thick,black] (5.5,2.5) -- (6.5,2.5);
\draw[thick,black] (5.5,3.0) -- (6.5,3.0);
\filldraw[blue!70] (6,1) circle (4pt);
\filldraw[blue!60] (6,2) circle (4pt);
\filldraw[blue!40] (6,1.5) circle (4pt);
\filldraw[blue!80] (6,2.5) circle (4pt);
\filldraw[blue!50] (6,0.5) circle (4pt);
\filldraw[blue!10] (6,3.) circle (4pt);
\filldraw[blue!20] (6,0.) circle (4pt);
\draw[very thick,->] (5.3,3.3) .. controls (7.25,4) .. (9.25,3.25);
\node at (7.3,4.2) {\textcolor{green!60!black}{\bf Truncation}};
\draw[thick,black] (8,0) -- (9,0);
\draw[thick,black] (8,0.5) -- (9,0.5);
\draw[thick,black] (8,1) -- (9,1);
\draw[thick,black] (8,1.5) -- (9,1.5);
\draw[thick,black] (8,2.0) -- (9,2.0);
\draw[thick,black] (8,2.5) -- (9,2.5);
\draw[thick,black] (8,3.0) -- (9,3.0);
\filldraw[red!80] (8.5,1) circle (4pt);
\filldraw[red!30] (8.5,2) circle (4pt);
\filldraw[red!60] (8.5,1.5) circle (4pt);
\filldraw[red!50] (8.5,2.5) circle (4pt);
\filldraw[red!70] (8.5,0.5) circle (4pt);
\filldraw[red!10] (8.5,3.) circle (4pt);
\filldraw[red!20] (8.5,0.) circle (4pt);
\draw[thick,black] (9.5,0) -- (10.5,0);
\draw[thick,black] (9.5,0.5) -- (10.5,0.5);
\draw[thick,black] (9.5,1) -- (10.5,1);
\draw[thick,black] (9.5,1.5) -- (10.5,1.5);
\draw[thick,black] (9.5,2.0) -- (10.5,2.0);
\draw[thick,black] (9.5,2.5) -- (10.5,2.5);
\draw[thick,black] (9.5,3.0) -- (10.5,3.0);
\filldraw[blue!70] (10,1) circle (4pt);
\filldraw[blue!60] (10,2) circle (4pt);
\filldraw[blue!40] (10,1.5) circle (4pt);
\filldraw[blue!80] (10,2.5) circle (4pt);
\filldraw[blue!50] (10,0.5) circle (4pt);
\filldraw[blue!10] (10,3.) circle (4pt);
\filldraw[blue!20] (10,0.) circle (4pt);
\draw [fill=green,opacity=0.4] (7.75,0.75) rectangle (10.75,2.25);
\draw [very thick,green!60!black] (7.75,0.75) rectangle (10.75,2.25);
\node at (1.5,-1) {$\ket{\psi_0^{(m)}}$};
\node at (5.5,-1) {$\ket{\psi^{(m)}}$};
\node at (9.5,-1) {$\ket{\psi^{(m)}_s}$};
\end{tikzpicture}
\caption{Graphical illustration of our model: Two $n-$level systems $(n = 7)$ are initialized in the $m\times m-$dimensional $(m=2)$ maximally entangled state \eref{input}. Subsequently, an entanglement preserving dynamics \eref{Spreading} populates all levels of both subsystems. Finally, the system's state is truncated (see equation \eref{truncated}) into the $s\times s-$dimensional subspace $(s=3)$ represented by the green rectangle.}
\label{fig:model}
\end{figure}
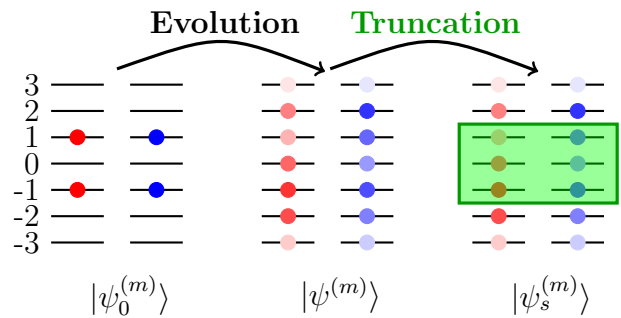

To systematically investigate this effect, we consider two $n-$level systems initially prepared in a maximally entangled state of an $m\times m-$dimensional subspace of their total Hilbert space. 
We then propagate this initial state by an entanglement preserving unitary operator that populates all $n$ levels in each factor space.
Upon projection onto a finite-dimensional, $s\times s$ subspace, we quantify the concomitant changes of the output state's entanglement (see figure \ref{fig:model}).
For general local (i.e. acting separately on the two subsystems) quantum dynamics, we find a simple expression for the output state entanglement, given as a function of the dimensions of the encoding, the total, and the truncation Hilbert spaces.

\section{Model}
We consider a bipartite quantum system living in the Hilbert space $\mathcal{H}=\mathcal{H}^n_A\otimes \mathcal{H}^n_B$, with $\mathcal{H}^n_A$ and $\mathcal{H}^n_B$ discrete Hilbert spaces of dimension $n = 2N+1$ ($N=1,2,\ldots$) each. 
We assume that the initial state of the total system is maximally entangled \cite{bennett1993} in the subspace $\mathcal{H}^m_A\otimes \mathcal{H}^m_B$, with local dimension $m =2M< n$ ($m =2M+1\leq n$) for even (odd) {\it encoding subspace}, where $M=1,2,\ldots, N^\prime$ and $N^\prime \leq N$. Explicitly, the initial state reads
\begin{equation}
\ket{\psi^{(m)}_0} = \frac{1}{\sqrt{m}}\left(\sum_{k=-M}^M\ket{k}_A\ket{-k}_B-f_m\ket{0}_A\ket{0}_B\right),
\label{input}
\end{equation}
where $\ket{k}_{A/B}$, with $k\in \{-N, \cdots, N\}$, is a orthonormal basis of $\mathcal{H}^n_{A/B}$ and $f_m = [(-1)^m +1]/2$. Thus, for $m=2M$ the state $\ket{0}_A\ket{0}_B$ is excluded from the quantum superposition in equation \eref{input}, while it is included for $m=2M+1$.
A graphical illustration of this state is presented (for $m=2$) in the left column of figure \ref{fig:model}.

Application of local unitary operations to both parties leads to (see the central column in figure \ref{fig:model})
\begin{equation}
\ket{\psi^{(m)}} = U^{(A)}\otimes  U^{(B)}\ket{\psi^{(m)}_0}=\sum_{q,r = -N}^N\beta_{q,r}\ket{q}_A\ket{r}_B,
\label{Spreading}
\end{equation}
with 
\begin{equation}
\beta_{q,r} = \frac{1}{\sqrt{m}}\left(\sum_{k=-M}^{M}U^{(A)}_{q,k}U^{(B)}_{r,-k} - f_m U^{(A)}_{q,0}U^{(B)}_{r,0} \right).
\label{beta}
\end{equation} 

In the next step, we project $\ket{\psi^{(m)}}$ onto a subspace $\mathcal{H}^s_A\otimes \mathcal{H}^s_B$ of $\mathcal{H}$, where both factor spaces have the  dimension $s=2S+1$ ($S=1,2,\ldots, N^{\prime \prime}$, with $N^{\prime \prime} \leq N$).  
This procedure is accomplished with the help of a truncation operator $T_s$, such that 
\begin{equation}
\ket{\psi^{(m)}_s} = T_s\ket{\psi^{(m)}} = \sum_{q,r = -S}^S\beta^\prime_{q,r}\ket{q}_A\ket{r}_B,
\label{truncated}
\end{equation}
where $\beta^\prime_{q,r} = \beta_{q,r}/\mathcal{N}$ (with $\mathcal{N}^2 = \sum_{q,r = -S}^S | \beta_{q,r}|^2$ ) ensures that $\ket{\psi^{(m)}_s}$ has unit norm.
This last step is illustrated (for $s=3$) in the right column of figure \ref{fig:model}.

Since $\ket{\psi^{(m)}_s}$ is a pure state, we can quantify its entanglement using the purity of the reduced density matrix $\mathcal{P}=\tr\rho_{A}^2$ \cite{MintertReview2005}, with 
\begin{equation}
\rho_{A} = \tr_B[\ket{\psi^{(m)}_s}\bra{\psi^{(m)}_s}] = \sum_{q, r, k  = -S}^{S} \beta^\prime_{q, r} \beta^{\prime *}_{k, r} (\ket{q}\bra{k})_A.
\label{reducedrho}
\end{equation}
More precisely, we employ the Schmidt number $K = 1/\mathcal{P}$ \cite{GrobeJPB1994,LawPRL2004}, which ranges from $1$ for a separable state to $d$ for a $d\times d-$dimensional maximally entangled state. 
\added{Using equation \eqref{reducedrho}}, the purity of the reduced density matrix, which is also related to other entanglement measures like concurrence \cite{WoottersPRL1998,RungtaPRA2001}, can be expressed in terms of the coefficients $\beta^\prime_{q,r}$ as 
\begin{equation}
\mathcal{P} = \sum_{q,r,k,l = -S}^S \beta^\prime_{q,r}\beta^{\prime*}_{k,r} \beta^\prime_{k,l}\beta^{\prime*}_{q,l}.
\label{RedPurity}
\end{equation}

Because a local unitary cannot affect the entanglement of $\ket{\psi^{(m)}_0}$ \cite{horodecki2009,MintertReview2005},
the only effect of the transformation \eref{Spreading} is to modify the basis representation of the entanglement inscribed into the state, possibly spreading it over the entire Hilbert space $\mathcal{H}$.  
In contrast, the truncation $T_s$ is a {\it local non-unitary} operation that preserves entanglement if and only if the latter is confined in the truncation subspace. In all other cases, the operator $T_s$ induces entanglement losses.
We now set out to quantify these losses for different dimensions $m$ and $s$, and for different choices of the unitary operators $U^{(A)}$, $U^{(B)}$.

\section{Uniform spreading} 
We start by considering unitary operations that transform any state of the computational basis $\ket{k}_{A/B}$ into an equally weighted superposition of all basis states. This is accomplished if we choose $U^{(A)} = U^{(B)} = U$ as
\begin{equation}
U\ket{k} = \ket{\alpha_k} = \frac{1}{\sqrt{n}}\sum_{l=-N}^{N}e^{i\frac{k l}{n}2\pi}\ket{l},
\label{MUB}
\end{equation}
where the phase factors in equation \eref{MUB} ensure that the states $\ket{\alpha_k}$ form a complete orthonormal basis of $\mathcal{H}_{A/B}$. 
The sets $\{\ket{k}\}$ and $\{\ket{\alpha_k}\}$ are two mutually unbiased bases (MUB), meaning that an arbitrary state of either basis set is equally distributed over all the elements of the other \cite{Schwinger570,durt2010mutually}. 
We therefore refer to $U$ in equation \eref{MUB} as the {\it uniform spreading} operator.

\begin{figure}
\centering
\includegraphics[width=\columnwidth]{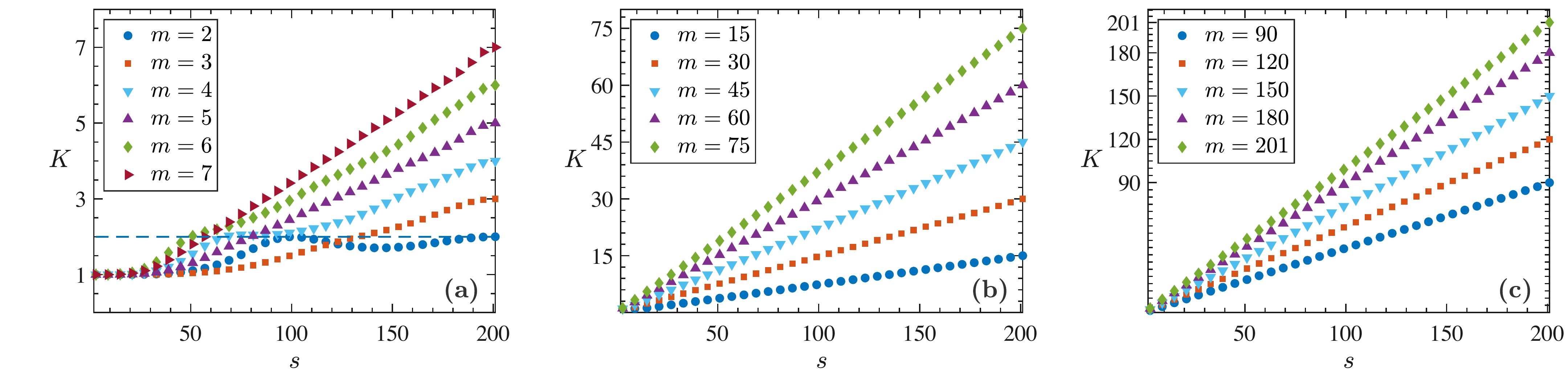}
\caption{(color online) 
Schmidt number $K = 1/\mathcal{P}$ of the truncated state $\ket{\psi_s^{(m)}}$ \eref{truncated}, obtained from different $m\times m-$dimensional 
maximally entangled states $\ket{\psi_0^{(m)}}$ \eref{input} evolved via the transformation \eref{Spreading} with $U^{(A)}$ and $U^{(B)}$ given by the uniform spreading operator \eref{MUB}, plotted against the truncation dimension $s = 3, 5, \ldots, n = 201$. The dashed horizontal line in (a) represents the amount of entanglement of the initial state \eref{input} with $m=2$.}
\label{Fig:UniformSpreading}
\end{figure}

Substituting the matrix elements of the latter for $U^{(A)}$ and $U^{(B)}$ into equation \eref{beta}, we obtain the coefficients
\begin{equation}
\beta_{q,r} = \frac{(m+f_m)\mathrm{sinc}\left[(q-r)(m+f_m)\pi/n\right]}{n\sqrt{m}\;\mathrm{sinc}\left[(q-r)\pi/n\right]}  -\frac{ f_m}{n\sqrt{m}}
\label{betaMUB}
\end{equation}
to determine the purity \eref{RedPurity} of the reduced state.
In general, the sum in \eref{RedPurity} cannot be calculated analytically, but is easily assessed numerically. 
Figure \ref{Fig:UniformSpreading} shows an exemplary case of the dependence of $K$ on the truncation dimension $s = 3, 5, \ldots, 2N+1$, with $N =100$.

Let us first discuss the scenario where each subsystem is initialized in a two-dimensional subspace, obtained by setting $m=2$ ($M=1$) in \eref{input}. 
In this case, the Schmidt number of the truncated-state depends non-monotonically on the truncation dimension $s$. 
Moreover, for this particular initial state, equation \eref{betaMUB} simplifies to $\beta_{q,r} = \sqrt{2}\cos[2\pi(q-r)/n]/n$, which allows for an analytic summation in equation  \eref{RedPurity}, with the result
\begin{equation}
\mathcal{P}_{m=2} = \frac{1}{2} + \frac{8s^2\sin^2(2\pi /n)\sin^2(2\pi s/n)}{\left[s^2 \cos(4\pi/n) +\cos (4\pi s/n) - s^2 -1 \right]^2}.
\label{P2}
\end{equation}
We notice that, in order to recover the entanglement of the initial state ($K = 1/\mathcal{P} = 2$), the second term in equation \eref{P2} must vanish. 
This condition is satisfied for $s = n, n/2$. 
The first solution is obvious: No entanglement is lost in the absence of truncation. 
The condition $s=n/2$ cannot be realized exactly, since, by construction, both $n$ and $s$ are odd integers. 
Yet, closer inspection of the data plotted in figure  \ref{Fig:UniformSpreading} (a) shows that for $s\approx n/2$ the entanglement of the truncated state differs from that of the input state by only a fraction of a per cent.
Similar modulations are observed for $m=4,6$ and, very weakly, for $m=3$ in figure \ref{Fig:UniformSpreading}  (a), but in all these cases the amplitudes of the secondary maxima or shoulders are much smaller than the untruncated states' entanglement. 
\added{These modulations come from the here considered particular unitary [see equation \eqref{MUB}] for which $\mathcal{P}$ corresponding to small $m$ is a sum of few oscillating terms given by equation \eqref{betaMUB}. }

For all values of $m$ except $m=2,4$, the Schmidt number of the truncated state grows monotonically with the truncation dimension $s$. 
If $m \ll n$, there are two almost flat regions around the extrema (at $s=3$ and $s=n$), which are connected by an effectively linear growth [see, for example, $m=5,6,7$ in figure \ref{Fig:UniformSpreading} (a) ].
With increasing $m$, the size of the flat regions shrinks and $K$ behaves essentially linearly [figure \ref{Fig:UniformSpreading} (b - c)]. 
This linear behaviour is very well approximated by $K = ms/n$, 
which converts into an exact expression for $m=n$. 
In fact, the input state $\ket{\psi^{(m=n)}_0}$  is invariant under the transformation \eref{MUB} in this case and $T_s\ket{\psi_0^{(n)}} = \ket{\psi_s^{(n)}} = \ket{\psi_0^{(s)}}$, which is the maximally entangled state in dimension $s$, giving $K = s$.

\section{Random unitary}

After discussing how the entanglement of a uniformly spread $m-$dimensional maximally entangled state is affected by truncation, we now want to understand how this, rather special, case compares to a general local transformation. 
To this end, we consider local random unitaries from the \textit{circular unitary ensemble} (CUE) \cite{mehta}, i.e. the group of $n\times n$ unitary matrices $U(n)$ uniformly distributed according to the Haar measure \cite{conway1994,Mezzadri2005}.

\begin{figure}[t]
\centering
\includegraphics[width=\columnwidth]{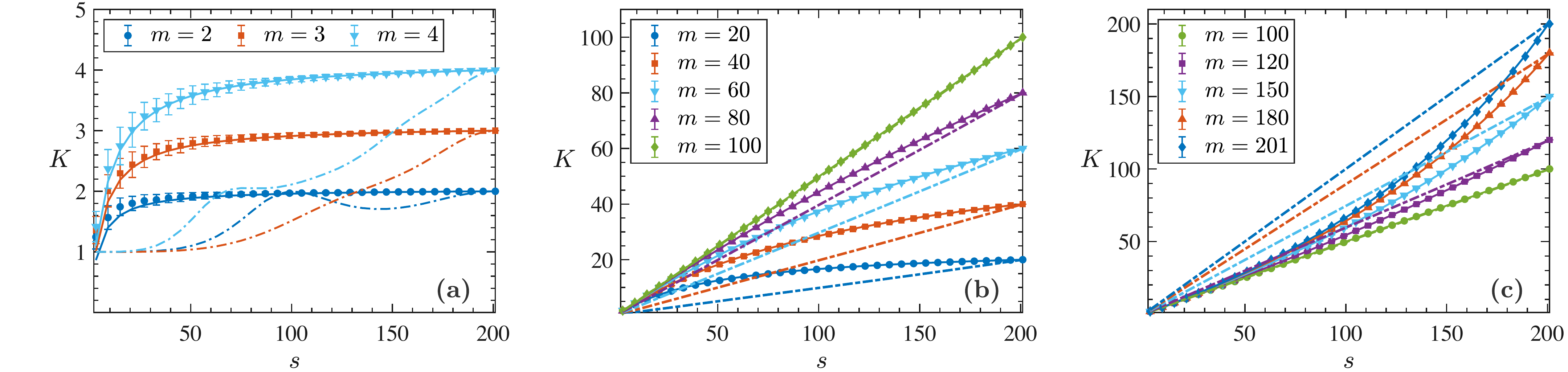}
\caption{(color online) Mean Schmidt number $K = 1/\mathcal{P}$ of the truncated state $\ket{\psi_s^{(m)}}$ \eref{truncated}, obtained from different $m\times m-$dimensional maximally entangled states $\ket{\psi_0^{(m)}}$ \eref{input} evolved via the transformation \eref{Spreading} with $U^{(A)}\neq U^{(B)}$ independent random unitaries, plotted against the truncation dimension $s = 3, 5, \ldots, n = 201$. Solid lines represent equation \eref{fit}, while symbols with error bars were obtained extracting the mean value and the standard deviation of $K$ from $100$ random realizations of  $U^{(A)}$ and $U^{(B)}$. The results for uniform spreading \eref{MUB} (dot-dashed lines of the corresponding color) are included for comparison. }
\label{Fig:RandomSpreading}
\end{figure}

For each pair of random unitaries $U^{(A)} \neq U^{(B)}$ in equation \eref{Spreading}, we calculate the output entanglement for different dimensions $m =2,\ldots, n = 201$ and $s= 3,\ldots, 201$ of the initial state and of the truncation subspace, respectively. 
We finally extract the mean value and the standard deviation of the Schmidt number $K = 1/\mathcal{P}$, which are plotted in figure \ref{Fig:RandomSpreading}, for $100$ independent random realizations of $U^{(A)}$ and $U^{(B)}$.

A first noteworthy feature of figure \ref{Fig:RandomSpreading} is   the size of the error bars decreasing with $m$ and $s$. 
This observation can be understood by invoking the concentration of measure phenomenon \cite{ledoux2001,milman2009}, 
according to which any `well behaved' function on a hypersphere concentrates around its mean value 
\footnote{More precisely, for real-valued Lipschitz continuous functions \cite{ledoux2001,milman2009} on an $n-$dimensional hypersphere, deviations from the mean (evaluated with respect to the uniform distribution on the hypersphere) larger than $\epsilon$ are exponentially suppressed both in $\epsilon$ and $n$.}. 
The set of pure states forms a hypersphere in $\mathcal{H}$ \cite{bengtsson2017} and entanglement is well behaved thereon \cite{Tiersch2013,Hayden2006, Lubkin1978, Benenti2009}.
Therefore, since $\ket{\psi_0^{(m)}}$ is pure and both the transformation \eref{Spreading} and the truncation \eref{truncated} does not affect the state's purity, the concentration of measure phenomenon ensures that most local unitary transformations produce the same entanglement behaviour when combined with truncation.

Looking at the numerical results (symbols with error bars in figure \ref{Fig:RandomSpreading}), we were able to conjecture an expression for the reduced purity of a truncated state
\begin{equation}
\mathcal{P} = \frac{2}{s} + \frac{1}{m} -\frac{2}{n}.
\label{fit}
\end{equation}
Despite its simplicity, equation \eref{fit} (solid lines in figure \ref{Fig:RandomSpreading}) shows a very good agreement with the numerical data, especially for higher-dimensional input states, while for lower-dimensional encoding subspaces ($m\lesssim 5$) it slightly underestimates the data in the small $s$ region. equation \eref{fit} is the main result of this work.

From  equation \eref{fit}, we see that [figure \ref{Fig:RandomSpreading} (a)], in contrast to our observation for the uniform spreading case above, even for small values of the initial-state dimension $m$ the entanglement of $\ket{\psi_s^{(m)}}$ increases monotonically with $s$.
Moreover, the mean of $K$ is a concave function, with values, except for $m=2$ and $s \approx n/2$, larger than those obtained for the uniform spreading (reproduced for comparison as dot-dashed lines in figure \ref{Fig:RandomSpreading}).

For $s\ll n,m$, the first term in equation \eref{fit} is dominant resulting in a linear behaviour of the Schmidt number $K = 1/\mathcal{P} \approx s/2$. 
Therefore, all lines in figure \ref{Fig:RandomSpreading} [see in particular panels (b) and (c)] start out with the same slope.
With increasing values of $s$, the curves corresponding to different values of $m$ deflect either upwards (for $m\gtrsim n/2$) or downwards  (for $m\lesssim n/2$) from the $s/2$ line to reach the value $K = 1/\mathcal{P} = m$ at $s=n$. 
Consequently, the Schmidt number \eref{fit} is concave for $m < n/2$ and convex for $m > n/2$.
This change of convexity implies that the uniform spreading studied above gives a lower bound of entanglement in all truncated subspaces, for all $m \lesssim n/2$ [figure \ref{Fig:RandomSpreading} (b)]; conversely, it yields an upper bound for $m \gtrsim n/2$  [figure \ref{Fig:RandomSpreading} (c)]. 
For $m = n/2$, equation \eref{fit} reduces to $K = 1/\mathcal{P} = s/2$, hence the Schmidt number of a truncated state that underwent random unitary evolution or uniform spreading exhibits the same linear dependence on $s$.
\begin{figure}[t!]
\centering
\includegraphics[width=0.5\columnwidth]{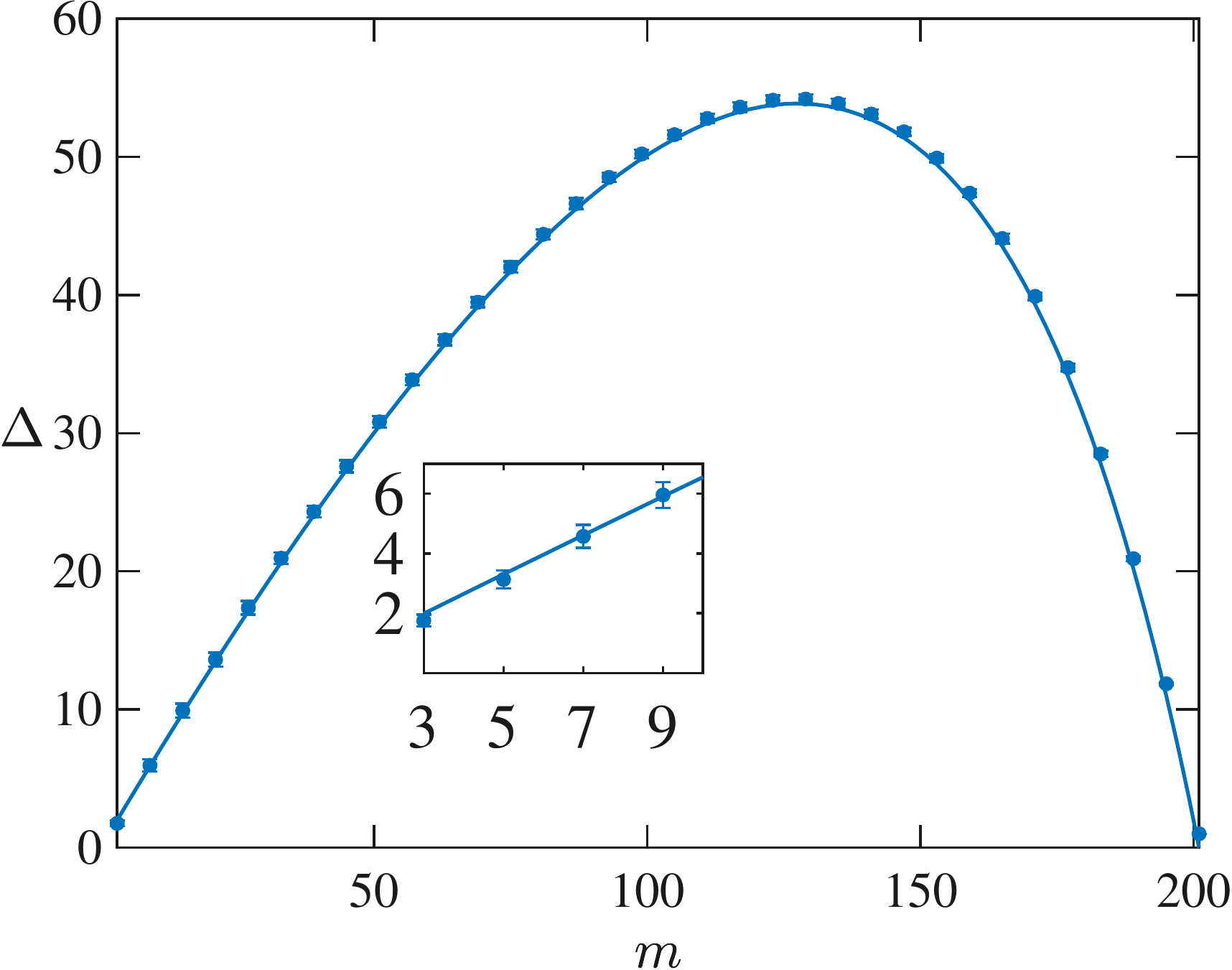}
\caption{(color online) Entanglement loss induced by truncation into the encoding subspace \added{($s=m=3,5,\dots, n=201$)} of states \eref{input} evolved according to \eref{Spreading} with $U^{(A)}$ and $U^{(B)}$ two random unitaries form the circular unitary ensemble. Symbols with error bars are numerical results, while the solid line represents equation \eref{Delta}.}
\label{Fig:EntLoss}
\end{figure}

To clarify the physical meaning of equation \eref{fit}, we now focus on truncation into the encoding subspace  ($s=m$). 
This case is particularly relevant in a quantum communication scenario, where one is only interested in reading the information that remains in the encoding subspace \cite{Groblacher_2006}. 
Let us therefore define the entanglement loss $\Delta$ as the difference between the entanglement of the initial state \eref{input} and the one of the truncated state \eref{truncated}. 
With equation \eref{fit}, this reads
\begin{equation}
\Delta = m - K|_{s=m} = \frac{2 m (m-n)}{2 m-3 n}.
\label{Delta}
\end{equation}
equation \eref{Delta} is plotted (solid line), together with the corresponding numerical results (symbols with error bars), in figure \ref{Fig:EntLoss}. 
We notice that, for $m < (2-\sqrt{3})n$, increasing the dimension of the encoding subspace increases the entanglement losses due to truncation. 
Since in many practical applications \cite{Mair2001,Franke-Arnold2002,Dada2011}, the dimension of the encoding subspace is in general much smaller than the total Hilbert space dimension ($m \ll n$), our results (see inset in figure \ref{Fig:EntLoss}) predict a linear growth of the truncation-induced entanglement loss with the dimension of the encoding states. 
Evidence of such behaviour was reported in the context of free space quantum communication across a turbulent atmosphere \cite{Zhang:2016,Sorelli_2019} .
There entanglement of input states -- encoded in orbital angular momentum (OAM) states of light with $m= 2,3,4$-- decayed the faster the larger $m$, at a given turbulence strength.
While input states were strongly localized in the OAM basis, turbulence-induced crosstalk (mediated by local unitaries as presently considered) led to a spreading of the transmitted state beyond the encoding subspace. 
Projection on the latter ultimately led to the observed entanglement loss.

\section{Conclusion}
We investigated how the entanglement of a high-dimensional bipartite quantum system is affected by truncation into a subspace of its total Hilbert space.

Studying uniformly distributed random unitary matrices we showed that local unitary transformations produce the same entanglement losses when combined with truncation. 
Moreover, we provided a simple and accurate analytical approximation for the truncation-induced entanglement loss. 
This approximation predicts, in the experimentally most relevant case of an encoding subspace much smaller than the total Hilbert space, an enhanced entanglement loss with increasing encoding dimension.
This immediately applies to the experimentally relevant setting of entanglement transport across atmospheric turbulence \added{where truncation is the main cause of entanglement loss}.

The main difference between the model considered here and the free-space transmission of photonic entanglement is that we only considered dynamics that does not affect the state's purity. \added{Conversely, atmospheric turbulence would mix the state.}
In the future, it will be useful to investigate how truncation affects mixed states.
As a further matter, the degrees of freedom available in photonic systems are either two-dimensional (polarization) or infinite dimensional (spatial modes \cite{Mair2001,Franke-Arnold2002,Dada2011}, frequency \cite{REICHERT1999,Roslund2013}, time \cite{Thew:2004,Brendel:1999}), forcing one to truncate the Hilbert space in practical implementations. 
It will thus be worthy to face the theoretical challenge of studying truncation in infinite dimensional Hilbert spaces.

\ack
{GS thanks Mattia Walschaers and Manuel Gessner for illuminating and enjoyable discussions on the use of random unitary matrices. G.S., V.N.S. and A.B. acknowledge support by Deutsche Forschungsgemeinschaft under grant DFG BU 1337/17-1.

\section*{References}
\bibliography{Biblio}
\bibliographystyle{iopart-num}
\end{document}